\documentclass{PoS}

\def\fig#1{Fig.~\ref{#1}}

\def\gev{\mbox{~GeV}}

\def\gevc{\mbox{~GeV/$c$}}

\def\la{\left< }
\def\ra{\right> }

\def\meankv#1{\ensuremath{\la#1^2\ra}}

\def\sqrtrms#1{\ensuremath{\sqrt{\meankv{#1}}}}

\newcommand{\pp} {\ensuremath{p+p}}

\def\ptq#1{\ensuremath{\hat{p}_{T\rm #1}}}

\def\pt#1{\ensuremath{p_{T\rm #1}}} 
 
\def\vpt#1{\ensuremath{\vec{p}_{T\rm #1}}}

\def\kt#1{\ensuremath{k_{T\rm #1}}}

\newcommand{\ptt}{\ensuremath{p_{T\rm t}}}

\newcommand{\zt} {\ensuremath{z_{\rm t}}}

\newcommand{\za} {\ensuremath{z_{\rm a}}}

\newcommand{\xe} {\ensuremath{x_{\rm E}}}

\newcommand{\s} {\ensuremath{\sqrt{s}}}
\newcommand{\piz} {\ensuremath{\pi^0}}

\def\bgi{\begin{itemize}}
\def\endi{\end{itemize}}
\def\bge{\begin{equation}}
\def\ende{\end{equation}}
\def\bgc{\begin{center}}
\def\endc{\end{center}}

\title{Jet properties from direct $\gamma$ - hadron correlation in PHENIX at RHIC}

\ShortTitle{Jet properties from direct $\gamma$ - hadron correlation in PHENIX at RHIC}

\author{\speaker{D.J Kim}, Sami Räsänen, Jan Rak\\
        Jyväskylä University, Finland\\
        E-mail: \email{djkim@jyu.fi}}

\usepackage{subfigure}

\abstract{Two-particle correlations of direct photon triggers
with associated hadrons are obtained by isolation method in \pp\ collisions at \s = 200 \gev\ in PHENIX at RHIC. 
The initial momentum of the away-side parton is tightly constrained, because the parton-photon pair is balanced in momentum at the leading order in perturbative quantum chromodynamics (pQCD). Therefore making such correlations can be used as a tool to measure the away-side parton fragmentation function. The direct photon associated yields in \pp\ collisions are compared with PYTHIA~\cite{sjostrand-2006-026}
and the effect of the $k_{T}$ smearing in the spectra is discussed.}

\FullConference{4th international workshop High-pT physics at LHC 09\\
		 February 4-7, 2009\\
		 Prague, Czech Republic}
		 
\begin{document}
\section{Introduction}
\label{sec:gammajet}
Fragmentation functions represent the probability for a parton to fragment into a particular hadron carrying a certain fraction of the parton's energy. Fragmentation functions incorporate the long distance, non-perturbative physics of the hadronization process in which the observed hadrons are formed from final state partons of the hard scattering process. It, like structure functions, cannot be calculated in perturbative QCD but can be evolved from a starting distribution at a defined energy scale~\cite{Kniehl:2000hk}. If the fragmentation functions are combined with the cross sections for the inclusive production of each parton type in the given physical process, predictions can be made for the scaled momentum, $z$, spectra of final state hadrons. Small $z$ fragmentation is significantly affected by the coherence (destructive interference) of soft gluons~\cite{pQCDbook1991}, while scaling violation of the fragmentation function at large $z$ allows to measure $\alpha_{s}$~\cite{Abreu1997194}.
It has been believed that the shape of the high-\pt{}\ trigger hadron associated \xe\ distributions reflects the shape of the
fragmentation function~\cite{PhysRevD.15.2590}, by \xe\ we mean
\bge\label{eq:xe}
\xe=-{\vpt{t}\cdot\vpt{a}\over\pt{t}^2}=-{{\pt{a}\over\pt{t}} {\cos\Delta\phi}}
\simeq {\za\,\ptq{a}\over\zt\,\ptq{t}}
\ende
where \ptq{t}, \ptq{a}, \pt{t}, \pt{a} are the transverse momenta of the
trigger and associated parton, trigger and associated hadrons and
\za=\pt{a}/\ptq{a} and \zt=\pt{t}/\ptq{t}.
However, it has been demonstrated in~\cite{longPRC:2006sc} that
the fixed momentum of the trigger particle \pt{t}\ does not fix the mother parton
momentum \ptq{t}\ and \pt{t}\ varies with \pt{a}. In this case both \za\ and
\zt\ vary with \pt{a}\ and this variation completely masks the actual shape of the
fragmentation function.
One of the alternative ways to explore the fragmentation function is to study the particle
distributions associated to the direct photon. In the high-\pt{}\ region where the
photon-production is dominated by the Compton scattering, the direct photon
balances the back-to-back quark. The associated \xe\ distribution then approximates the fragmentation function of the away side jet.
However, in this case there are important effects
which need to be taken into account. One constraint comes from the fact that at low \pt{}\
the direct photon production may be contaminated by fragmentation photons from the large
number of gluonic jets produced ~\cite{Arleo:2007qw,Vitev:2008vk}. Fragmentation photons
obviously do not balance the away-side parton.
Another constraint comes from the fact that even when considering the leading order Compton
diagram there is always soft QCD radiation which breaks the jet-photon momentum balance and the
azimuthal collinearity. This non-perturbative radiation manifests itself as a non-zero value of the net parton-photon pair transverse momentum magnitude. 
Originally this soft radiation induced transverse momentum was attributed to the individual incoming partons
and the notation of \kt{}\ (parton transverse momentum) was introduced by Feynman, Field and Fox \cite{Feynman4}, where the colliding partons might have some initial transverse momentum \kt{} with respect to the incoming hadrons. This analog of the Fermi motion of nucleons in a nucleus would give
rise to a smearing out of the \pt{} spectrum. There could be an additional smearing due to the \kt{} associated with the fragmentation in the final state \cite{RevModPhys.59.465}.

\section{PHENIX $\piz\ -h$ and direct $\gamma - h$ correlation results}

\begin{figure}[ht]
\begin{center} 
\resizebox{7.5cm}{!}{\includegraphics{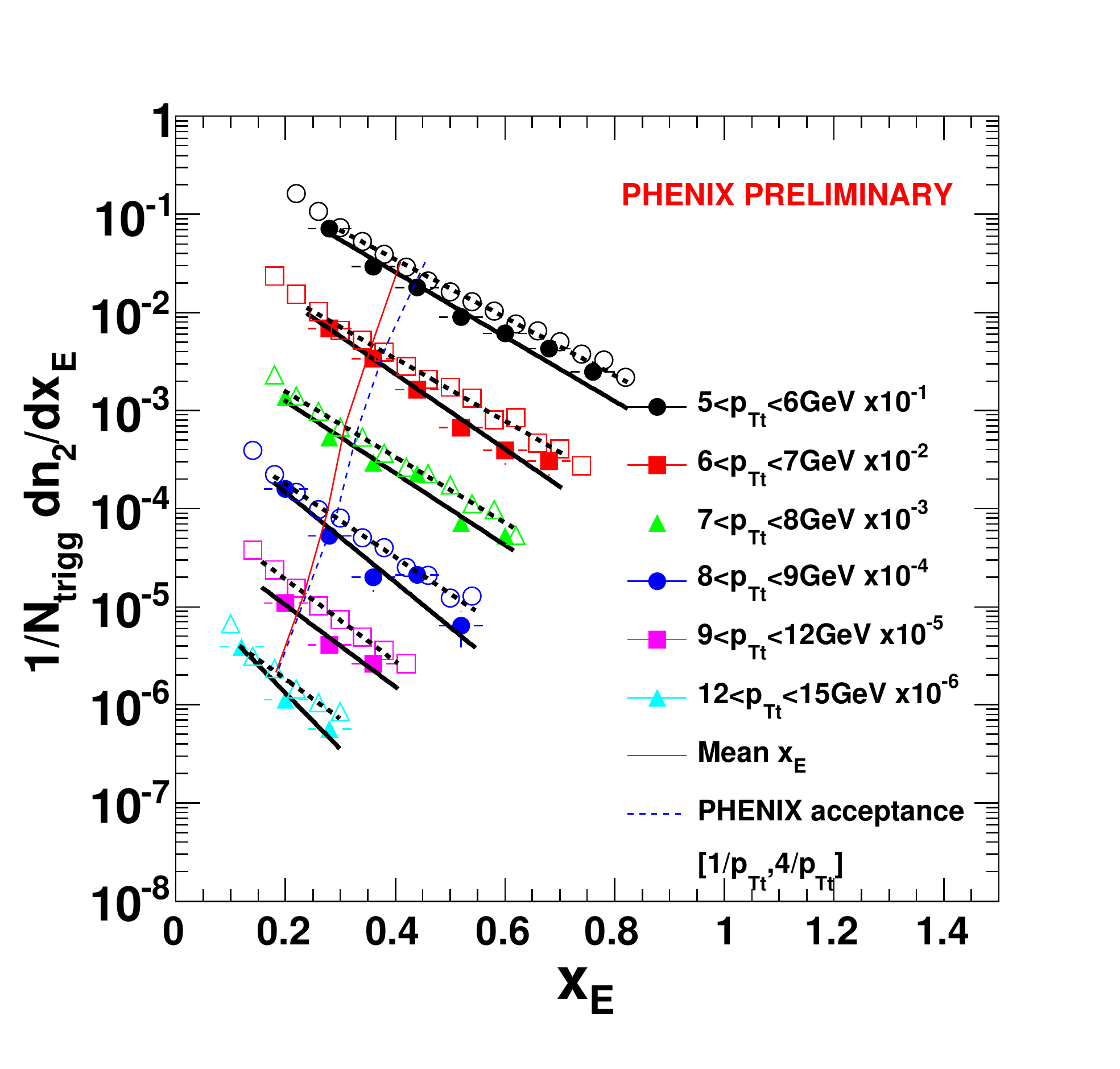}}
\resizebox{7.5cm}{!}{\includegraphics[angle=90]{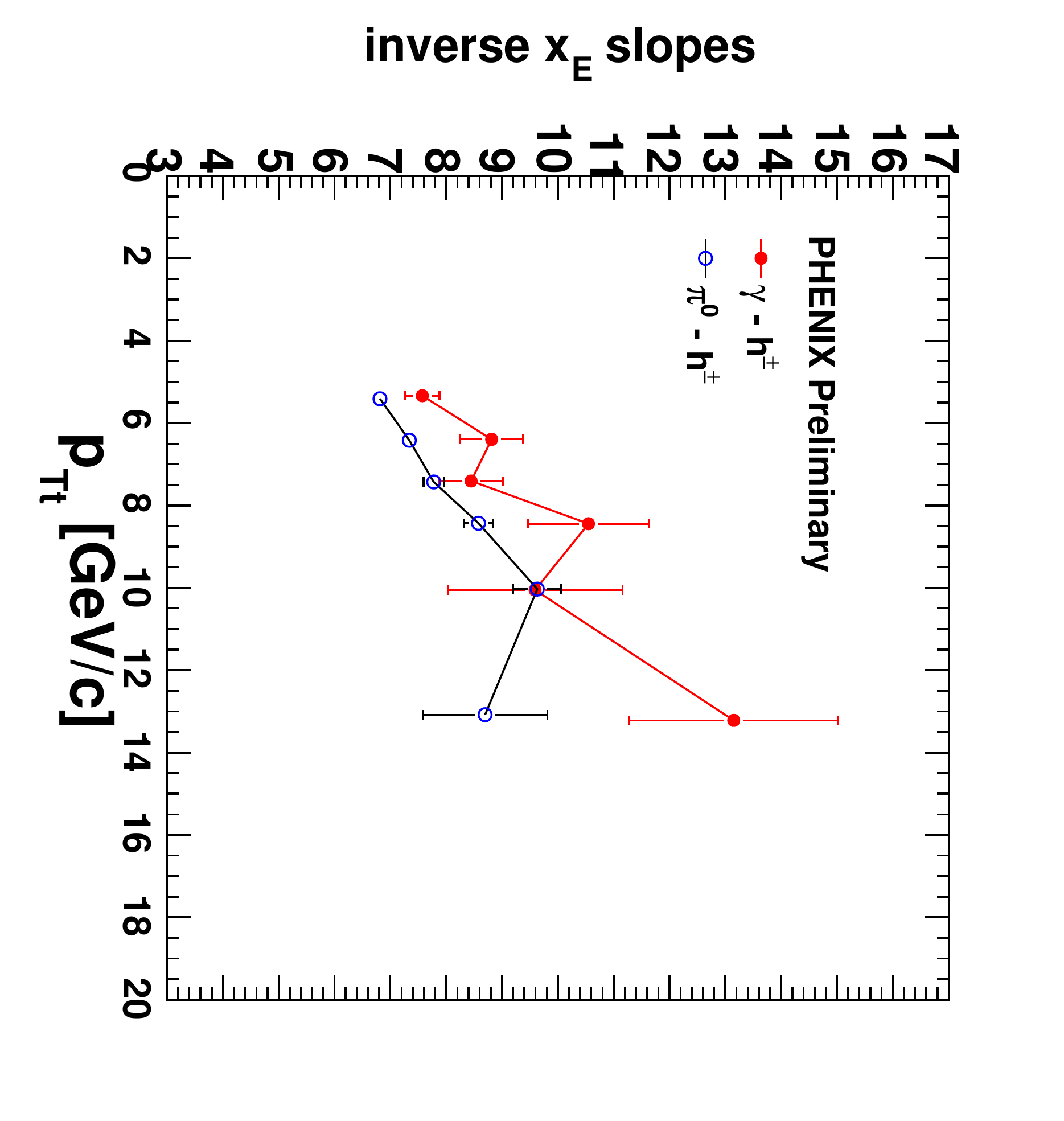}}
\caption{
 PHENIX  preliminary data from \pp\ collisions at \s =200 \gev. Left : \xe\ distributions of  charged hadrons associated with trigger $\piz$ (open markers)  and isolated $\gamma$ (closed markers) are fitted to a simple exponential function ($N e^{-b\xe}$) in the available range of \xe. The mean \xe\ values extracted from the distributions are indicated by a solid line which goes across the \xe\ distributions labeled as "Mean \xe". As for the dotted line, labeled as "PHENIX acceptance", the mean \xe\ is calculated by limiting the associated charged hadron momenta from 1 to 4 \gevc\ of PHENIX acceptance in a given trigger momentum. 
Right : The extracted exponential slope $b$ (called "inverse \xe\ slope" in the figure) is shown as a function of  trigger particle momentum.}
\label{fig:gammarun6pp} 
\end{center}
\end{figure}

The identification of direct photons is difficult due to the large number of background
photons from hadronic decays, mostly from \piz\ decays. Therefore the extraction of the direct photon-hadron pairs per trigger yields relies on a statistical subtraction of the decay photon-hadron per trigger yields from the inclusive photon-hadron pairs per trigger yields. The first result from this "subtraction method" was presented in ~\cite{Nguyen:2008ic}.
Recently PHENIX developed photon isolation cuts 
which were applied event by event in a new analysis in order to dramatically reduce the 
contamination of di-jet events with \piz\ decay or fragmentation photons in the $\gamma-h$ 
event sample. 
The measured \xe\ distributions of charged hadrons associated with $\pi^0$ and
isolated photon triggers in \pp\ collisions at $\s=200$ GeV are presented in \fig{fig:gammarun6pp} for various isolated photon and $\pi^0$  trigger momenta.
A fit of the \xe\ distribution to a simple exponential ($N e^{-b\xe}$)  in the available range of \xe\ is presented in left panel of the \fig{fig:gammarun6pp}. The extracted exponential slope $b$ from the fitting as a function of trigger \pt{t}\ is 
shown in the right side of the \fig{fig:gammarun6pp}.
A very interesting feature from this result is observed. The \xe\ slopes are still increasing as the trigger momentum gets larger, which is very similar to the $\pi^{0}-h$ case.

\begin{figure}[htp]
  \begin{center}  
    \subfigure[PYTHIA quark fragmentation function]{\label{fig:xe_quark_pythia200}\includegraphics[scale=0.37]{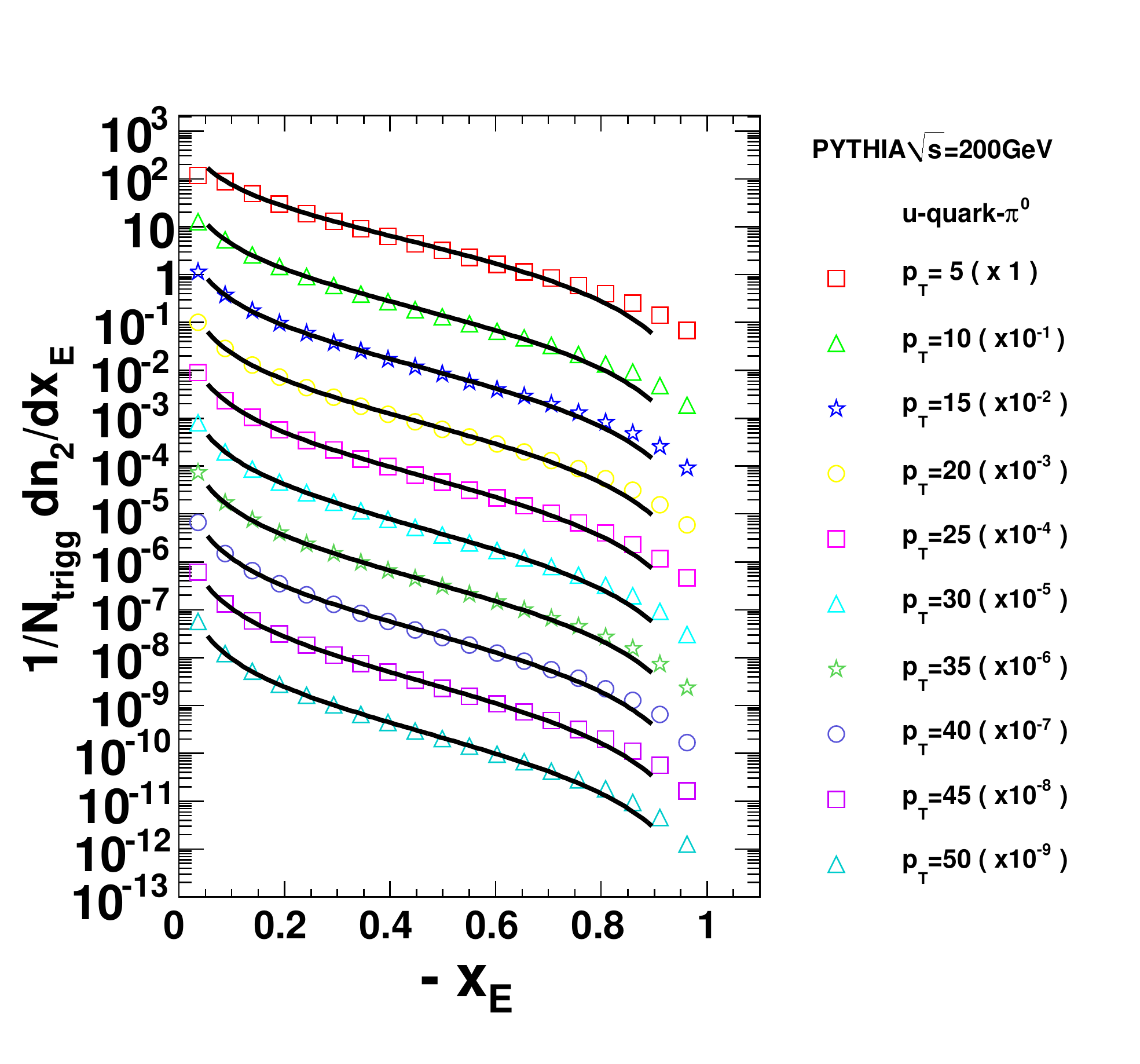}}
    \subfigure[PYTHIA gluon fragmentation function]{\label{fig:xe_gluon_pythia200}\includegraphics[scale=0.37]{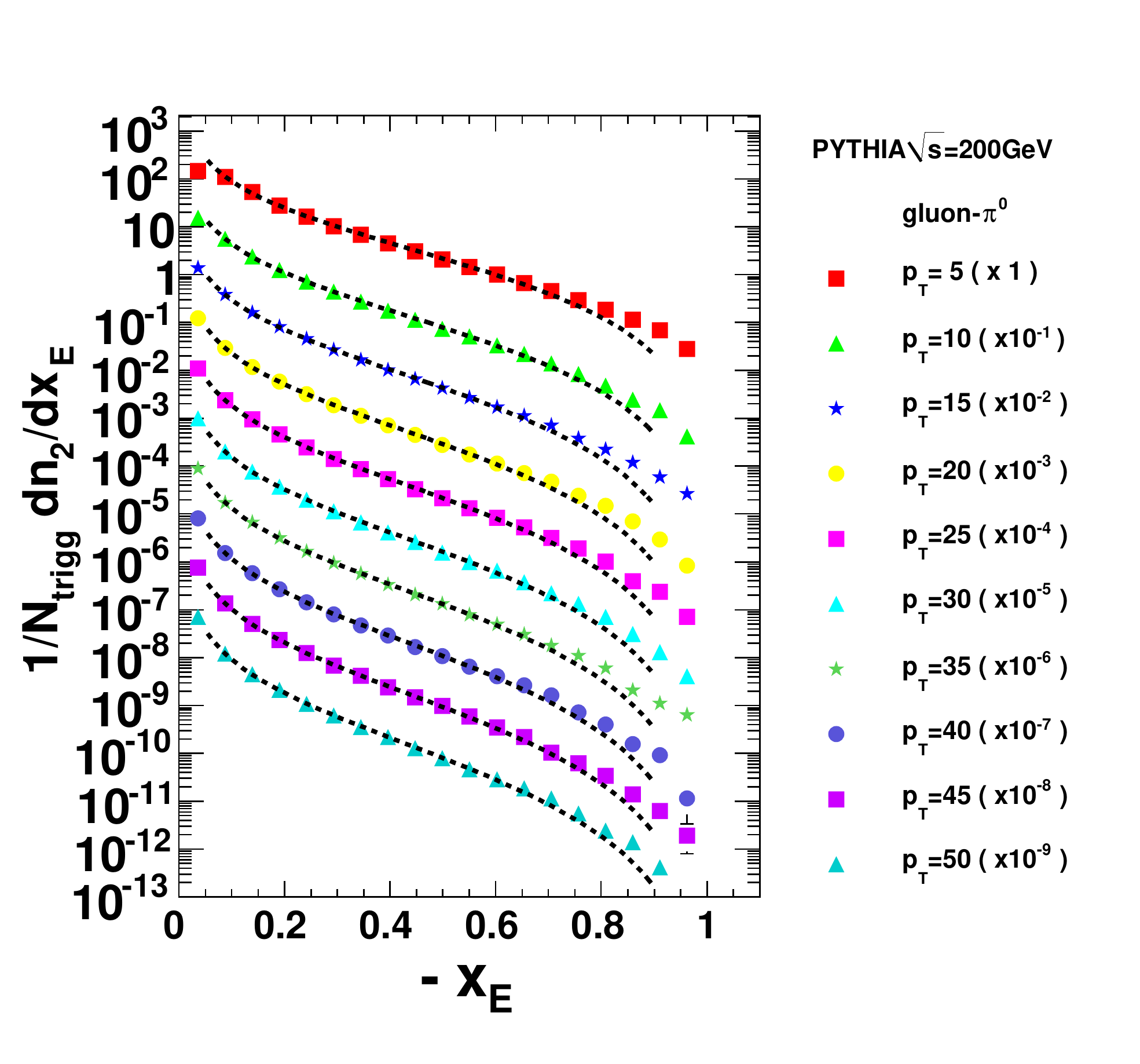}}
    \end{center}
\caption{ \xe\ distributions of \piz's resulting from quark and gluon fragmentation in various quark/gluon momenta, \ptt{}\  in PYTHIA (\pp\ at \s =200 \gev). The results of the KKP parametrization~\cite{Kniehl:2000hk} fitting to the distributions are shown as solid (for u-quarks)
and dashed (for gluons) lines. The momenta in the figure label are given in \gevc.}
\label{fig:pythia200_xe}
\end{figure}
In order to understand this behavior of \xe\  slopes as a function of trigger momentum, we compare the PHENIX preliminary data with PYTHIA, where  $\sqrtrms{\kt{}} = 3.0 \gevc$ is used based on the PHENIX measurement~\cite{longPRC:2006sc}.
\fig{fig:pythia200_xe} shows the \xe\  distributions of quark and gluon fragmenting to \piz's for various quark and gluon momenta calculated by PYTHIA (note that \xe\ is negative because only near side particles associated with quarks and gluons are accounted for).
The \xe\ distributions of \piz's resulting from quark and gluon fragmentation are parameterized by means of a fit function
\bge\label{eq:fit}
\frac{dN}{d\xe}\propto \xe^{-\alpha}(1-\xe)^\beta,
\ende
i.e. we use in fitting the same functional form as in KKP parametrization~\cite{Kniehl:2000hk} 
for fragmentation functions. The results of the fitting are shown as solid (for u-quarks)
and dashed (for gluons) lines.
\begin{figure}[htp]
  \begin{center}  
    \subfigure[quarks]{\label{fig:nld_quark_pythia200}\includegraphics[scale=0.33]{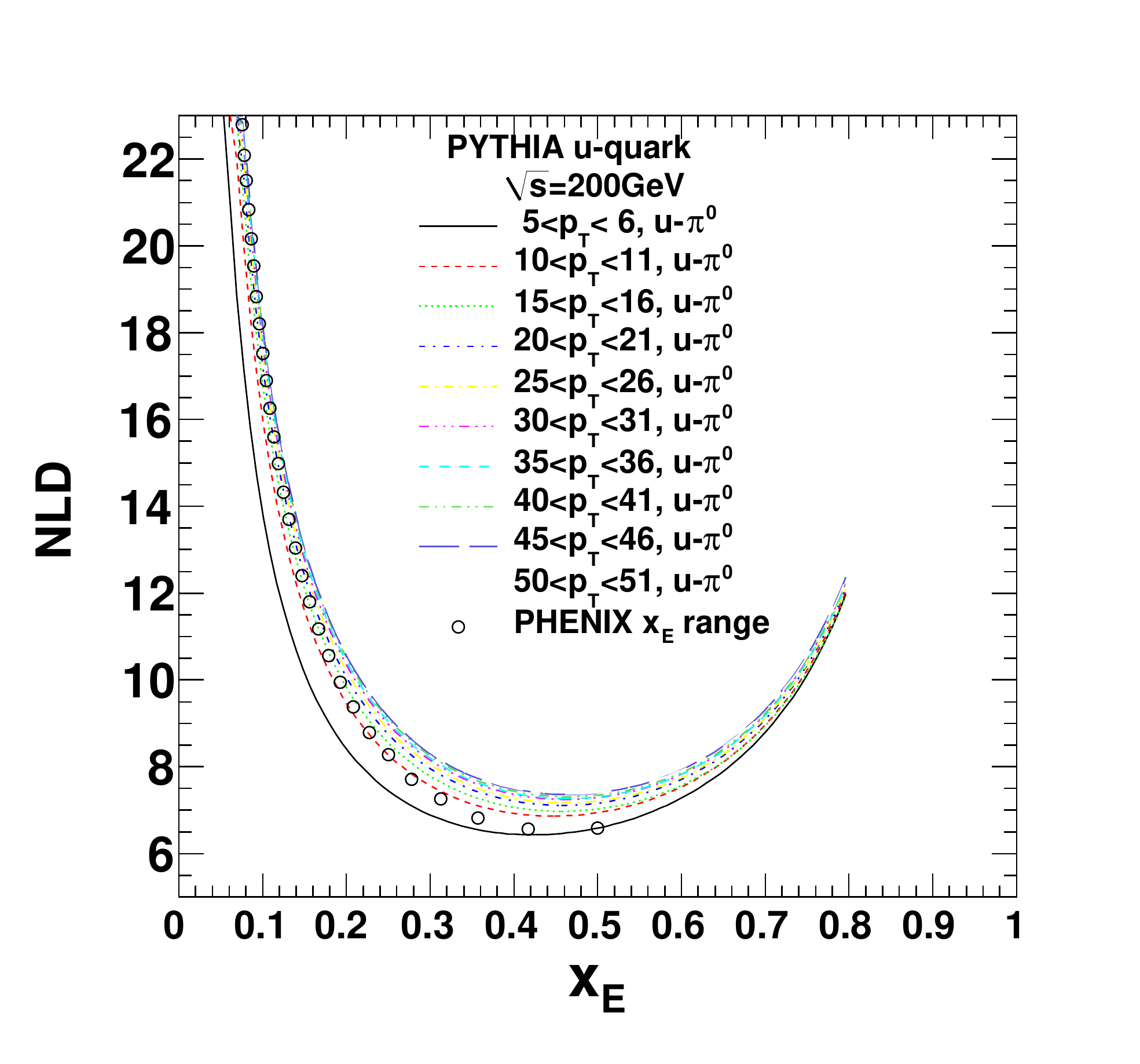}}
    \subfigure[gluons]{\label{fig:nld_gluon_pythia200}\includegraphics[scale=0.33]{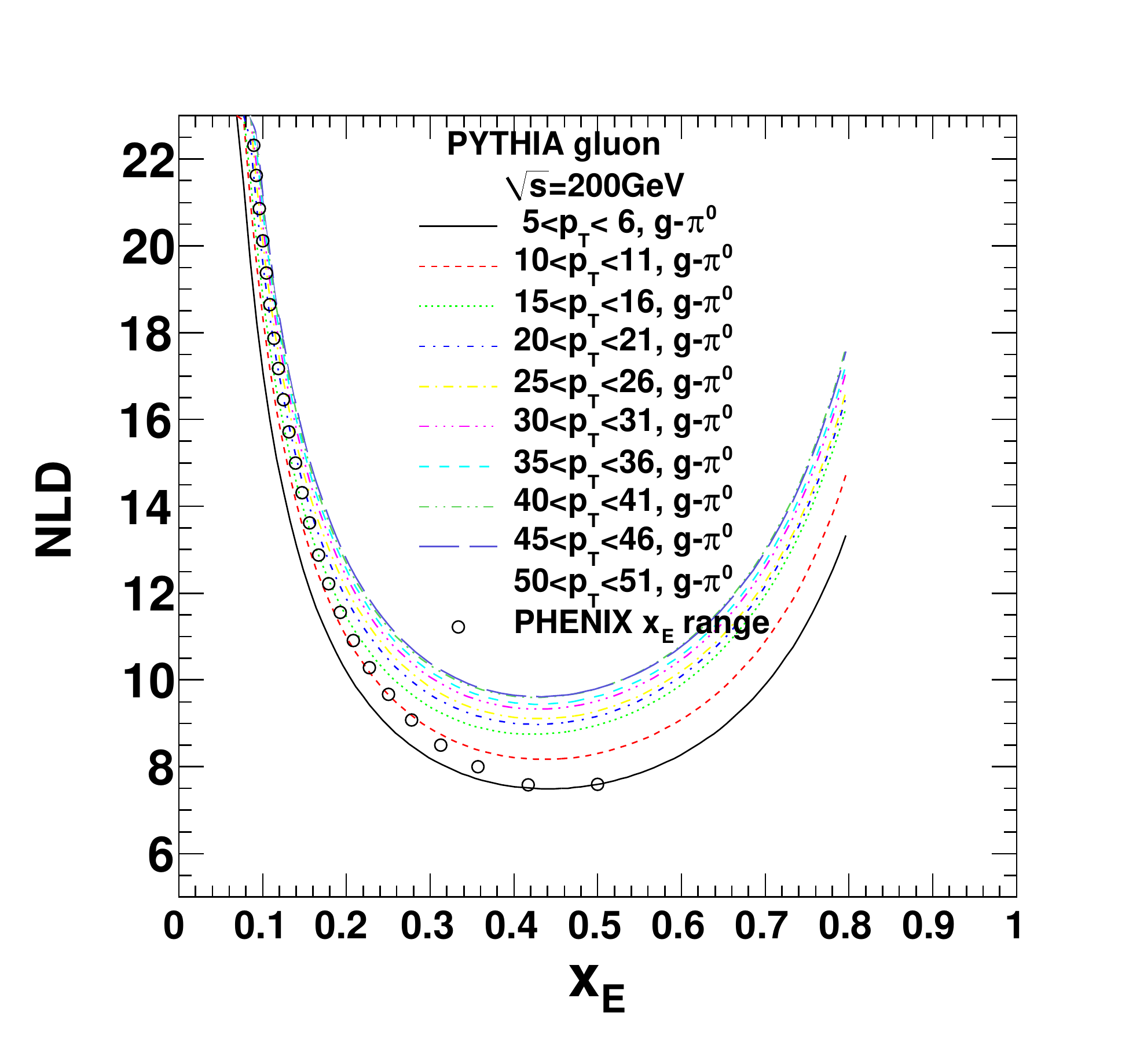}}
    \end{center}
\caption{PYTHIA  \pp\ at \s =200 \gev. Negative Logarithmic Derivative ($NLD = - \frac{d}{d\xe} \ln(\frac{dN}{d\xe})$) of  the  \xe\ distributions of \piz's resulting from quarks and gluons fragmentation in various quark/gluon momenta. As for the open circle  labeled as "PHENIX \xe\ range", the mean \xe\ is calculated by
 limiting the associated charged hadron momenta from 1 to 4 \gevc\ of PHENIX acceptance in the given trigger momentum.}
\label{fig:pythia200_nld}
\end{figure}
Negative Logarithmic Derivative (NLD) of \xe\ distributions
\bge\label{eq:NLD}
\mathrm{NLD}\equiv -\frac{d}{d\xe}\ln\left(\frac{dN}{d\xe}\right)
\ende
measures the local slope of the \xe\ distribution (NLD of an exponential
distribution is constant).
The NLD's extracted from the \xe\ distributions of
neutral pions associated with quarks and gluons as a function of \xe\ is presented in \fig{fig:pythia200_nld}.
Because the \xe\ distributions in the \xe\ range above $0.8$  cannot be properly described by the fit (see \fig{fig:pythia200_xe}), the NLD distributions are shown only below $\xe\ \approx 0.8$.
The NLD's for low \xe\ regions (less than $\xe\ \approx  0.2$) are much steeper than any other range, stay in lower values in the intermediate range ($ 0.2 \le \xe\ \le 0.7 $) and get larger for $\xe\  \ge 0.7$ for all quark and gluon trigger momenta. The NLD's from PYTHIA in the range of available PHENIX data are shown in the same \fig{fig:pythia200_nld} as open circles for each quark/gluon momentum. The mean \xe\ extracted from the PHENIX data and the \xe\ values used to extract "NLD" from PYTHIA by limiting the hadron acceptance of PHENIX are comparable to each other, as can be seen from \fig{fig:gammarun6pp}.
In the latter comparisons of the PHENIX data to PYTHIA, the mean \xe\ values labeled as "PHENIX acceptance" are used for higher momenta in PYTHIA where PHENIX data are not available.
\begin{figure}[htp]
  \begin{center}  
    \resizebox{7.00cm}{!}{\includegraphics{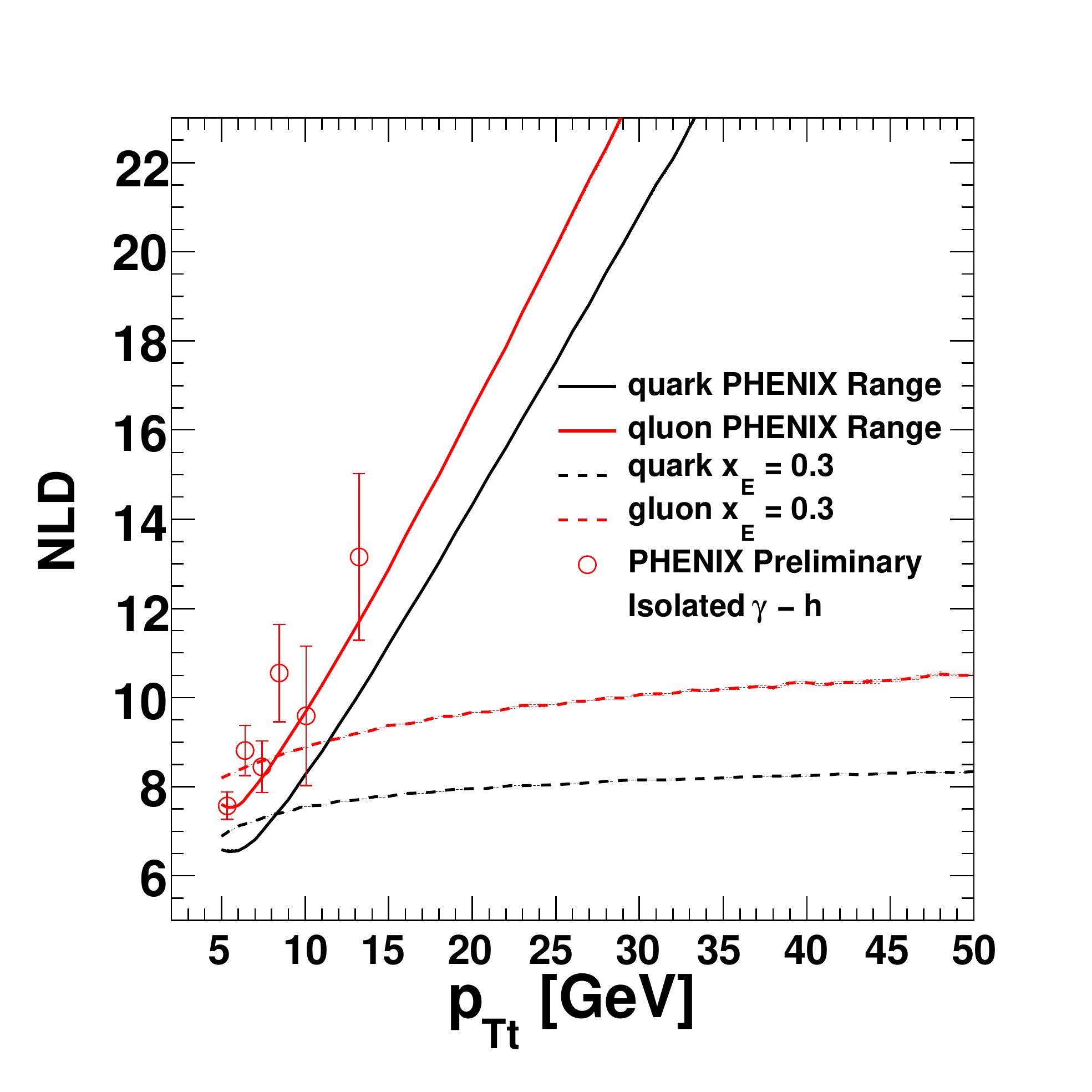}}
    \end{center}
 \caption{Inverse \xe\ slopes from PHENIX preliminary as a function of isolated $\gamma$  trigger momentum are compared with NLD's from \xe\ distributions of \piz's resulting from PYTHIA quark and gluon fragmentation with two set of \xe\ ranges. The solid lines  labeled as "PHENIX \xe\ range", represent the local slopes in the expected \xe\ value by considering PHENIX acceptance for charged hadron momentum in PYTHIA. The other shown as dotted line represents the NLD's for a fixed value of $\xe = 0.3$ from PYTHIA.}
\label{fig:nld_all}
\end{figure}

Finally, we have compared the exponential slopes of PHENIX data with PYTHIA where the used \xe\ values are close to the mean values of PHENIX \xe\ ranges in \fig{fig:nld_all}. They reveal the same increasing trend for the higher trigger \pt{t}\ and the local slopes of quarks and gluons from PYTHIA are comparable to PHENIX data. The steepness of the exponential slopes in PHENIX data is driven by the experimental limit of the \xe\ range, i.e away-side hadron can be measured only from 1 to 4\gevc\ in PHENIX, therefore
\xe\ value gets smaller as the trigger momenta gets higher.
As for the intermediate \xe\ ($\approx 0.3$), the absolute slopes are slightly increasing as the trigger momentum gets larger and tend to converge into a constant above 10 \gevc\  for quark and gluon fragmentation functions, shown as dotted lines. But the increment of the slopes for the gluon fragmentation functions (slope changes from 8 to 10) is rather larger about by factor of two than for the quark fragmentation functions (slope changes from  7 to 8) in the range $0.2 \le \xe\ \le 0.7$ (see \fig{fig:pythia200_nld}) when the trigger momentum varies from 5 to 50 \gevc.
The fact that the absolute NLD values in all trigger bins agree with those from quark and gluon fragmenation function suggests that the measured \xe\ distribution associated with direct photon is a representation of the quark or gluon fragmenation function. But a more detailed analysis with larger \pt{}\ trigger bins and wider \xe\ ranges is necessary to study trigger momentum and \xe\ dependent effects which can influence the fragmentation function.



\section{Conclusions}


The PHENIX experiment at the Relativistic Heavy Ion Collider (RHIC) has
measured \xe\ distributions of charged hadrons associated with $\pi^0$ and
isolated photon triggers in \pp\ collisions at $\s=200$ GeV. It has been
demonstrated in~\cite{longPRC:2006sc} that \xe\ distributions measured in di-hadron
correlations are insensitive to the fragmentation function. Direct photon
associated spectra are expected to follow fragmentation function more
closely, because in leading order two-to-two processes the trigger photon
fixes exactly the momentum of outgoing parton by momentum conservation.
However, several complications, like fragmentation photons and soft QCD
radiation, are expected. Nevertheless, the measured data presents a
striking feature that slopes of the \xe\ distributions, in both
$\pi^0$ and photon trigger cases, continue to rise as a function of trigger
momentum.
We have studied this rising trend in the case of photon trigger by fitting
the same functional form as used in KKP parameterization~\cite{Kniehl:2000hk} to the \xe\
distributions of neutral pions resulting from fragmentation of u-quark
or gluon in PYTHIA. As the fit function is not exponential, we use
Negative Logarithmic Derivative (NLD) to measure local slopes. We found
that the observed rising trend of slopes of \xe\ distributions is driven
by the experimental limits of the \xe\ range. PHENIX can measure photons up
to very large momenta while for hadrons the measurable range is more
limited.
Therefore the measured \xe\ values for the higher trigger momenta are very small
where the \kt{} effect dominates and make the local slopes of the fragmentation function larger. 
Improvement of the statistical and systematic precision of the measurements with wider \xe\ ranges
and larger \pt{} trigger should allow further tests of fragmentation function as well as understanding of  \kt{} effect quantitatively.

\bibliographystyle{h-physrev3}
\bibliography{main_proc}
\end{document}